\documentclass{aa}
\usepackage{natbib,psfig,graphicx,ulem}
\usepackage{txfonts}
\usepackage{soul,color}

\def\mathnew{\mathsurround=0pt}
\def\simov#1#2{\lower .5pt\vbox{\baselineskip0pt \lineskip-.5pt
\ialign{$\mathnew#1\hfil##\hfil$\crcr#2\crcr\sim\crcr}}}

\def\MeV{Me\kern-0.11em V}
\def\keV{ke\kern-0.11em V}

\begin{document}

\title{The cluster Abell 780: an optical view \thanks{Based on
observations obtained at the European Southern Observatory, program ESO
68.A-0084(A), P.I. E.~Slezak. This research has made use of the
NASA/IPAC Extragalactic Database (NED) which is operated by the Jet
Propulsion Laboratory, California Institute of Technology, under
contract with the National Aeronautics and Space Administration.}}

\author{
F.~Durret \inst{1,2} \and
E.~Slezak \inst{3} \and
C.~Adami \inst{4} 
}

\institute{
UPMC Universit\'e Paris 06, UMR~7095, Institut d'Astrophysique de Paris, 
F-75014, Paris, France
\and
CNRS, UMR~7095, Institut d'Astrophysique de Paris, F-75014, Paris, France
\and
University of Nice Sophia Antipolis, CNRS, Observatoire de la C\^ote
d'Azur, B.P. 4229, 06304 Nice Cedex 4, France
\and
LAM, P\^ole de l'Etoile Site de Ch\^ateau-Gombert,
38 rue Fr\'ed\'eric Joliot-Curie,
13388 Marseille Cedex 13, France
}

\date{Accepted . Received ; Draft printed: \today}

\authorrunning{Durret et al.}

\titlerunning{Abell 780: an optical view}


\abstract
{The Abell 780 cluster, better known as the Hydra A cluster, has been
thouroughly analyzed in X-rays. However, little is known on its
optical properties.}
{We propose to derive the galaxy luminosity function (GLF) in this
apparently relaxed cluster, and to search for possible environmental
effects by comparing the GLFs in various regions, and by looking at the
galaxy distribution at large scale around Abell~780.}
{Our study is based on optical images obtained with the ESO 2.2m
telescope and WFI camera in the B and R bands, covering a total region
of 67.22$\times$32.94 arcmin$^2$, or 4.235$\times$2.075~Mpc$^2$ for a
cluster redshift of 0.0539.}
{In a region of 500~kpc radius around the cluster centre, the GLF in the
R band shows a double structure, with a broad and flat bright part 
and a flat faint end that can be fit by a power law with
an index $\alpha \sim -0.85\pm 0.12$ in the 20.25$\leq$R$\leq$21.75
interval. If we divide this 500~kpc radius region in North+South or
East+West halves, we find no clear difference between the GLFs in these
smaller regions.  No obvious large scale structure is apparent within
5~Mpc from the cluster, based on galaxy redshifts and magnitudes
collected from the NED database in a much larger region than that
covered by our data, suggesting that there is no major infall of
material in any preferential direction. However, the Serna-Gerbal
method reveals the presence of a gravitationally bound structure of
27 galaxies, which includes the cD, and of a more strongly gravitationally bound structure of 14 galaxies.}
{These optical results agree with the overall relaxed structure of Abell
780 previously derived from X-ray analyses.}


\maketitle

\section{Introduction}

Galaxy evolution is known to be influenced by environmental effects,
which are particularly strong in galaxy clusters, where the effects of
several physical processes such as ram pressure stripping, galaxy
harassment, or infall of field galaxies are commonly observed.  The
analysis of galaxy luminosity functions (GLF), and in particular of
their faint-end slopes in several wavebands is a good way to trace the
history of the faint galaxy population. As summarized for example in
Table~A.1 of Bou\'e et al. (2008), this slope can strongly vary
from one cluster to another, and it can also depend upon the depth of
the galaxy catalogue considered; while most studies give slopes in the
range $-0.9$ to $-1.5$, several find faint-end slopes can be as steep as
$-2.3$.  Possible reasons for such a range of values (cosmic
variance of the background counts, different mass buildup histories of
clusters, systematic errors in the data analysis) were discussed in
Bou\'e et al. (2008) and will not be repeated here (see references to
the literature in this paper). Note that the faint end slope can
also depend on the chosen filter, but there is no clear evidence that
the steepest slopes are always found in the same wavelength range.

Another interesting property is that, in a given cluster, the faint end
slope of the GLF has been observed to vary from one region to
another. For example, this effect has been observed in several clusters
such as Coma (Lobo et al. 1997, Beijersbergen et al. 2002, Adami et
al. 2007a) or Abell~496 (Bou\'e et al. 2008), and it was shown that the
faint-end slope gives indications on the cluster formation history
(see e.g. Adami et al. 2007a and references therein).

We present here the analysis of the cluster Abell~780, at a redshift
$z=0.0539$ (NED data base), which appears to be very relaxed and without
clear substructure (Burns et al. 1994). This cluster is better known as
Hydra~A, because its cD galaxy coincides with the strong wide angle tail
radio galaxy Hydra~A (Owen et al. 1992). Hydra~A is well known in
X-rays, in particular because it is X-ray luminous (David et
al. 1990 give ${\rm L_X(1.0\ Mpc)=4\ 10^{44}}$ erg~s$^{-1}$ in the
0.5-4.5~keV band for a redshift of 0.0522 and a Hubble constant
H$_0$=70~km~s$^{-1}$~Mpc$^{-1}$) and rather hot, in spite of being a
poor cluster of richness 0 (Batuski et al. 1991): the gas temperature is
about 4~keV (${\rm kT_X=4.5 \pm 0.7}$~keV from Einstein data, David et
al. 1990, ${\rm kT_X=3.80 \pm 0.12}$~keV from ROSAT data, Mohr et
al. 1999). At large scales, the thorough analysis of a deep
XMM-Newton image shows that the cluster indeed appears quite smooth and
relaxed (Simionescu et al. 2009a, 2009b). On the other hand, at much
smaller scales near the centre, deep Chandra data have revealed a number
of interesting features in this cluster: the hot gas in its innermost
regions shows cavities and a swiss cheese like topology, suggesting
matter outflow; evidence for the presence of a supermassive black hole
in the central cD galaxy has also been found (McNamara et al. 2000, Wise
et al. 2007).

On the other hand, very little has been published for this cluster at
optical wavelengths: two galaxy redshifts (Batuski et al. 1991), and
some imaging yet largely unpublished from the WINGS survey (Fasano et
al.  2006). No substructures were detected in Abell~780 from this survey
(Ramella et al. 2007). However, some galaxy redshift measurements are
available in NED. We present here optical images of Abell~780 in the B
and R bands covering a field of view of 67.22$\times$32.94 arcmin$^2$,
or 4.235$\times$2.075~Mpc$^2$.

The paper is organised as follows. We present our data and data
reduction in Section 2, and describe how the final galaxy catalogue is
obtained.  In Section 3, we describe the colour-magnitude relation and
GLFs obtained in the two bands, using the VVDS survey F02 field as a
comparison field for statistical subtraction of the foreground and
background galaxy populations (McCracken et al. 2003).  In Section 4,
based on galaxy redshifts found in the NED database, we analyze the
environment of Abell~780 at very large scale, and search for
gravitationally bound structures by applying the Serna-Gerbal method.

In order to facilitate comparisons with the X-ray results by Wise et
al. (2007) we will assume for Abell~780 a redshift $z=0.0539$, and a
flat $\Lambda$CDM cosmology with H$_0=70$~km~s$^{-1}$~Mpc$^{-1}$,
$\Omega _M=0.3$ and $\Omega _\Lambda =0.7$, yielding a luminosity
distance of 240~Mpc, a linear scale of 1.05 kpc~arcsec$^{-1}$ and a
distance modulus of 36.90.  In the absence of a direct estimate of the
galaxy velocity dispersion $\sigma _v$, we computed $\sigma _v$ from the
X-ray temperature measured by ROSAT applying the relation between
$\sigma _v$ and kT given by Girardi et al. (1996): $\sigma =
10^{2.59 \pm 0.04}{\rm T^{0.50\pm 0.07}}$. For kT=3.80~keV, we obtain
$\sigma _v$=758 km~s$^{-1}$. We then calculate the $r_{200}$ radius
using equation (8) from Carlberg et al. (1997) and find
$r_{200}$=1.91~Mpc.


\section{Data and methods}

\subsection{The data}

We observed the cluster Abell 780 with the ESO 2.2m telescope and the
WFI camera on March 18, 2002 (ESO Program 68.A-0084(A)).  Two fields
were observed, covering the East and West parts of the cluster
respectively, in both Johnson B and R filters (ESO filters B-842 and
R-844). 
The observations are summarized in Table~\ref{tab:obs}.  The camera
comprises 8 CCDs, giving a  field of about 8750 x 8305 pixels with
a pixel size of $0.238\times 0.238$ arcsec$^2$. Each WFI field is therefore
$34.708 \times 32.943$ arcmin$^2$, or $2187\times 2075$~kpc$^2$ at the
cluster distance.

\begin{figure} 
\centering \mbox{\psfig{figure=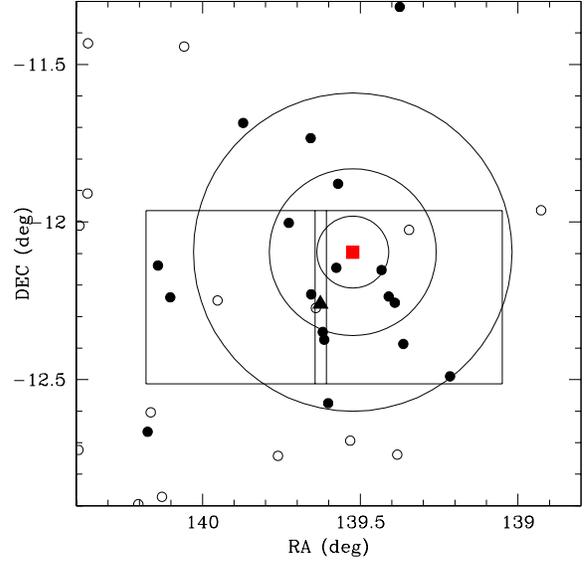,width=8cm}}
\caption[]{Map of the region of Abell~780. The two squares show the
positions of our images. The black triangle shows the Abell~780
cluster position according to the NED and Simbad databases, and the red
square that of the cluster centre as we define it (coinciding with the
Hydra A cD galaxy, see text). The three concentric circles show radii
of 0.5, 1.0 and 1.91~Mpc ($r_{200}$). Small circles show galaxies with
redshifts taken from NED, with filled circles indicating galaxies in the
[0.0338, 0.0738] redshift interval corresponding to a broad velocity
interval around the mean cluster velocity. } \label{fig:map}
\end{figure}

The two fields have a small area in common 2.19 arcmin wide in right
ascension (131.5 arcsec, or 552.5 pixels), where we believed that the
cluster centre was at the time of the observations (these were the
coordinates given for Abell 780 in the NED and Simbad databases).
However, we found out later that the position of the cD galaxy (the
Hydra A galaxy), which coincides with the maximum X-ray emission, is
offcentered by 6.1~arcmin in right ascension and 9.9~arcmin in
declination towards the North-West. We will take the position of the cD
as the cluster centre hereafter: $\alpha (2000.0) = 139.52375$~deg,
$\delta (2000.0) = -12.0955$~deg. Our mapping of the cluster therefore
misses part of its North half, but on the other hand allows us to probe
the South half of the cluster further away from the centre (see
Fig.~\ref{fig:map}). A larger fraction of the cluster is included in the
West field than in the East one. The total area covered by our two
fields is 67.22$\times$32.94 arcmin$^2$, or
4.235$\times$2.075~Mpc$^2$ in right ascension $\times$ declination.

\begin{table}
\caption{Summary of the observations (including coordinates of the two
 image centres).}
\begin{center}
\begin{tabular}{|l|ll|ll|}
\hline
 & East & & West & \\
\hline
RA (2000.0)  & 139.896093 &	& 139.354152 & \\
Dec (2000.0) & -12.2503	 &      & -12.25057  & \\
\hline
                & B     & R     & B     & R \\
Exp. time (s)   & 1500  & 1200  & 1500  & 1200 \\
Airmass	        & 1.084 & 1.049 & 1.138 & 1.216 \\
Seeing (arcsec) & 0.90  & 1.15  & 0.95  & 0.90 \\
Gain (e-/ADU)   & 2999  & 2399  & 2999  & 2399 \\
Zero Point      & 24.486 & 24.334 & 24.486 & 24.334 \\
ZP error   & $\pm 0.009$ & $\pm 0.005$ &  $\pm 0.009$ & $\pm 0.005$ \\
\hline
\end{tabular}
\end{center}
\label{tab:obs}
\end{table}

\subsection{Data reduction and source extraction}

The images were bias and flat field corrected in the usual way, using
flat fields obtained at twilight and dawn on the sky. For each field, 5
individual exposures were obtained with some dithering. They were
combined after being individually corrected for bias and flat field.  We
thus obtained a final set of four images in two bands, plus their
corresponding weight maps. These weight maps allow to take into
account the fact that due to the dithering applied during such
observations all the pixels in each image are not similarly exposed. The
SExtractor software takes them into account when detecting and measuring
objects on the images. Seeing measurements were made on unsaturated
stars on the four final stacked images. Since the East and West images
had slightly different seeings (at least in the R band), they were
processed independently (see Table~\ref{tab:obs}).

Photometric zero points were estimated from the observation of standard
stars from the Landolt list. 

Masks were made to exclude the regions with saturated pixels, 
bright stars and at the CCD edges. 


Object detection was performed by applying the SExtractor software
(Bertin \& Arnouts 1996). Detections were made in the R band (which is
more sensitive), then measurements were made in the B band as well with the
double image mode. After trying several values, we adopted a detection
threshold of 1.5~$\sigma$ and a minimum number of 5 pixels required for
an object to be detected. 

Total magnitudes (MAG-AUTO given by SExtractor) were computed in
the Vega system. Star catalogues were left in the Vega system in order
to be directly comparable with the Besan\c con counts (see Section
2.5).  Galaxy magnitudes were converted to the AB system in order for
the galaxy counts to be directly comparable to those of the VVDS (see
below). For this we applied the following
formulae:\footnote{http://www.astro.utoronto.ca/$\sim$patton/astro/mags.html\#conversions}
B(AB)=B(Vega)$-$0.163 and R(AB)=R(Vega)+0.055.

All magnitudes were corrected for galactic extinction, which is not negligible
since Abell~780 is at relatively low galactic latitude
(b=25$^\circ$). The Schlegel et al. (1998) extinction maps in the
direction of our two fields show the same mean extinction (within the
dispersion), with flux variations of only 11\% within each image,
corresponding to magnitude variations of at most 0.05 in B and 0.03 in
R, adopting a mean extinction value. We therefore adopted constant
extinction factors for both fields: 0.214 magnitudes in B and 0.132
magnitudes in R.

In view of the small redshift of Abell~780, we did not apply any
K-correction.

We thus obtained catalogues of objects in B and R for the East and West
fields separately.

\subsection{Catalogue completeness}

\begin{figure} \centering
\mbox{\psfig{figure=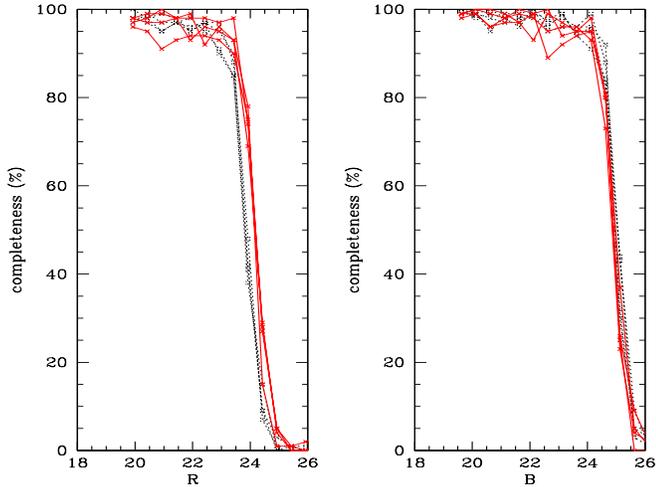,width=9cm,height=7cm,angle=0}}
\caption[]{Completeness in percentages in B (right) and R (left) for
point--like objects. The black dotted lines correspond to the East half
of the total field and the red full lines to the West half (where the
cluster is mainly located).}  \label{fig:compl}
\end{figure}

The completeness of the catalogue is estimated by simulations.  For
this, we add artificial stars of different magnitudes to the CCD images
and then attempt to recover them by running SExtractor again with the
same parameters used for object detection and classification on the
original images. In this way, the completeness is measured on the
original images and at different locations in the cluster.

In practice, we divide the full field of view in eight subimages, each
$4030\times 4030$~pixels$^2$ (similar to the subimage sizes already
adopted e.g. in Adami et al. 2006a). This represents a good
compromise between the spatial resolution of the completeness maps that
we compute and the poissonian error bars directly driven by the numbers
of objects considered in each subregion. Choosing eight subareas also
allows us to detect potential completeness variations from a WFI CCD to
another.  

In each subfield, and for each 0.5 magnitude bin between R=20 and 26,
we generate and add to the image one star that we then try to detect
with SExtractor, assuming the same parameters as previously. This
process is repeated 100 times for each of the eight subfields.

 The seeing variations between the East and West
images and in the R and B bands are taken into account. We also correct
magnitudes for extinction as for our object catalogue, and transform
them to AB magnitudes. Such simulations give a completeness percentage
for stars. This is obviously an upper limit for the completeness level,
since stars are easier to detect than galaxies. However, we have shown
in a previous paper that this method gives a good estimate of the
completeness for normal galaxies if we apply a shift of $\sim 0.5$~mag
(see Adami et al. 2006a). This does not apply to low surface
brightness galaxies, which are not expected to be detected here, mainly
because they would be fainter than our detection limit, as derived from
our results for LSBs in Coma (see Adami et al. 2006a and 2006b).
Results are shown in Fig.~\ref{fig:compl}.

From these simulations, and taking into account the fact that results
are worse by $\sim 0.5$~mag for mean galaxy populations than for stars,
we can consider that our galaxy catalogue is complete to better
than 90\% for R$\leq 23$ and 50\% for R$\leq 23.5$. In the B band, the
corresponding numbers for 90\% and 50\% completeness are B$\leq 23.7$
and B$\leq 24.3$ respectively.

\subsection{Star-galaxy separation}

\begin{figure} \centering
\mbox{\psfig{figure=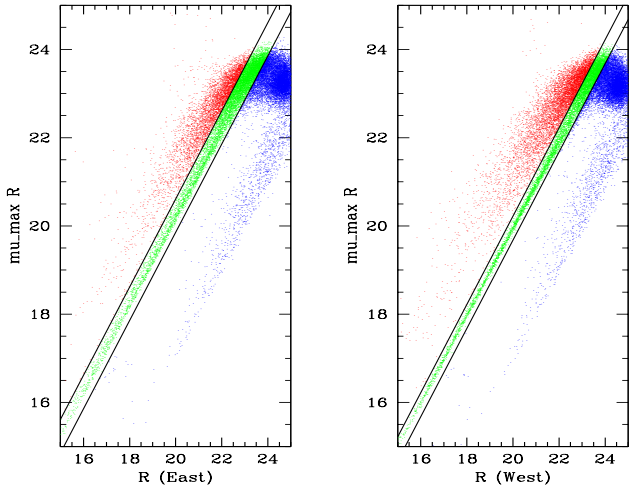,width=9cm,height=7cm,angle=0}}
\caption[]{Central surface brightness in the R band as a function of R
magnitude for the East (left) and West (right) fields. The lines
separate the galaxies (red points, above the top lines) from the stars
(green points, between the two lines). Objects below the bottom lines
(blue points) have been considered as defects. Objects with R$< 22$ and
$\mu _{\rm R}> 24$ are defects as well.}
\label{fig:sep_gal_star}
\end{figure}

In order to separate stars from galaxies, we plotted the central surface
brightness in the R band as a function of R magnitude for the East and
West fields (Fig.~\ref{fig:sep_gal_star}). Stars show well defined
sequences 
bound by lines with slopes of 1.00. 

In order to estimate the ordinate positions of the two upper lines
(separating stars from galaxies), we considered the star simulations
described in Sect.~2.3. for the R band. With respective seeings in the R
band of 1.15 and 0.90~arcsec in the East and West regions, the $1\sigma$
dispersions of the star distributions around a mean line of slope 1.00
are 0.127 and 0.091 respectively. We will consider that stars are
located at $\pm 3\sigma$ around the best line fit.  The corresponding
ordinate positions for R=15 of the two upper lines (those separating stars from
galaxies) are 15.61 and 15.22 for the East and West plots respectively.
All the objects above these lines have been classified as galaxies.

We looked individually at some objects located at the top edge of the
quoted sequence in the R=18-20 magnitude interval to check that
this separation was correct.


A few objects are visible in Fig.~\ref{fig:sep_gal_star} at (R$< 22$,
$\mu _{\rm R}> 24$); they do not correspond to low surface
brightness galaxies but were all checked to be tiny defects in the CCDs.

A number of ``objects'' appear to form a second, broader sequence below
the stars. On the images, these objects are found to be false detections
and are in a great majority located at the CCD edges or at the edges of
the masks drawn to cover bright stars. As seen in
Fig.~\ref{fig:sep_gal_star},  these defects can be separated from
stars by the two lower lines delimitating the star region, taken to be
symmetric of the star-galaxy separation line defined above relatively to
the line which best fits the star sequence.  The objects below these
two lower lines have been considered as defects and will be eliminated
from our catalogues hereafter.

This classification into stars, galaxies, and defects is of course not
valid over our whole range of magnitudes because of seeing
limitations, for example galaxy and star sequences appear to merge above
R$\sim$21.5 in Fig.~\ref{fig:sep_gal_star}.

We also checked by eye on the images that most of the 260 objects brighter
than R=18 and classified as galaxies were indeed galaxies and not
saturated stars.

\subsection{Star counts}

\begin{figure} 
\centering
\mbox{\psfig{figure=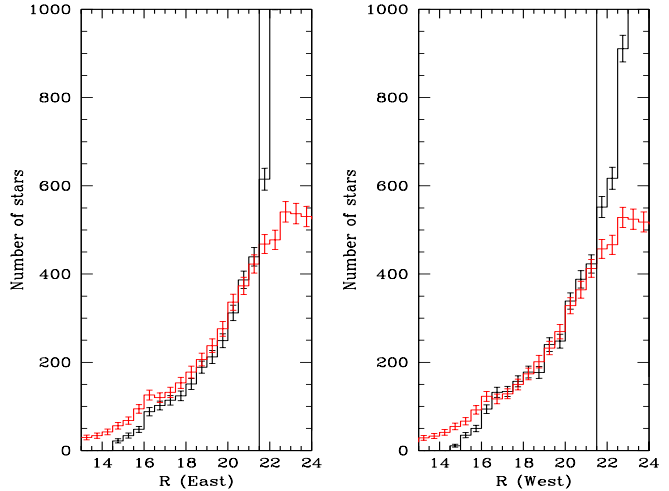,width=9cm,height=7cm,angle=0}}
\caption[]{Black histograms: star counts in the direction of Abell~780
in the East (left) and West (right) fields. The red histograms show the
star counts from the Besan\c con models normalized to the respective
surfaces of our fields. Error bars are Poissonian. The vertical lines at
R=21.5 show the limit beyond which our star-galaxy separation is not
reliable.}  \label{fig:histo_Bes_A780}
\end{figure}


In order to estimate more precisely the reliability of our star-galaxy
separation, we retrieved the star catalogue from the Besan\c{c}on model
for our Galaxy (Robin et al. 2003) in a 1~deg$^2$ region centered
on the position of Abell~780 given by NED (139.6265~deg, $-12.2611$~deg,
J2000.0). As seen from Fig.~\ref{fig:histo_Bes_A780}, the star count
histograms in the East and West regions estimated in Section 2.4 agree
with the Besan\c{c}on model star counts, very well in the West region
and within poissonian error bars for the East region, between magnitudes
R$\sim 17$ and R$\sim 21.5$.  At brighter magnitudes, the histograms do not
coincide, both because we masked very bright stars, and because our
total field is relatively small, leading to small number statistics for
bright stars.


At faint magnitudes (R$\geq 21.5$), our star counts steeply rise
while the Besan\c{c}on model counts increase much more smoothly. This
implies as expected that our star-galaxy separation is not reliable at
magnitudes larger than R=21.5 (or, to state this differently, the
$21.0-21.5$ bin is the faintest magnitude bin where the star-galaxy
separation can be considered as reliable).


\subsection{Galaxy counts}

In order to perform galaxy counts, we build a single catalogue of the
full East+West field after eliminating objects in common (we kept the
objects measured in the West field, since this is the field containing
a higher fraction of the cluster, and also the field with the best
seeing).  We consider that the star--galaxy separation described above
is reliable for R$<21.5$. Our full galaxy catalogue (with the East and
West regions merged) will be publicly available at the following
address: http://cencosw.oamp.fr/ .  It comprises 4256 galaxies brighter than
R=21.5.

For fainter magnitudes, we estimated the percentage of stars by
comparing the total number of detected objects (stars+galaxies) to
the predictions of the Besan\c con model in bins of 0.5~mag. We give
these percentages in Table~\ref{tab:pourcentstell}. The average 
fraction of galaxies was estimated in each magnitude bin from the
average of the star percentages.

\begin{table}
\caption{Percentage of stars in bins of 0.5~mag in the East and West
 fields, and average percentage of galaxies.}
\begin{center}
\begin{tabular}{|llll|}
\hline
R     & Stars & Stars & Galaxies \\
      & East  & West  & average \\
\hline
21.75 & 30.3\% & 29.6\% & 70.0\% \\
22.25 & 22.4\% & 23.9\% & 76.8\% \\
22.75 & 20.9\% & 21.8\% & 78.6\% \\
23.25 & 20.8\% & 19.5\% & 79.8\% \\
\hline
\end{tabular}
\end{center}
\label{tab:pourcentstell}
\end{table}

For R$\geq 21.5$ we will therefore consider that in each magnitude bin the
number of galaxies is equal to the total number of objects (stars +
galaxies) multiplied by the average percentage of galaxies. We will not
consider galaxy counts above R=23.5, since incompleteness becomes too
strong.

\begin{table}
\caption{Area effectively covered by each of the four regions of Fig.~\ref{fig:gal_counts}.}
\begin{center}
\begin{tabular}{|ll|}
\hline
Region & area \\
       & (deg$^2$) \\
\hline
r$<$500 kpc         & 0.049321 \\
r$<$1000 kpc        & 0.17635 \\
500$<$r$<$1000 kpc  & 0.12261 \\
1000$<$r$<$1500 kpc & 0.167136 \\
\hline
\end{tabular}
\end{center}
\label{tab:area}
\end{table}

\begin{figure} 
\centering
\mbox{\psfig{figure=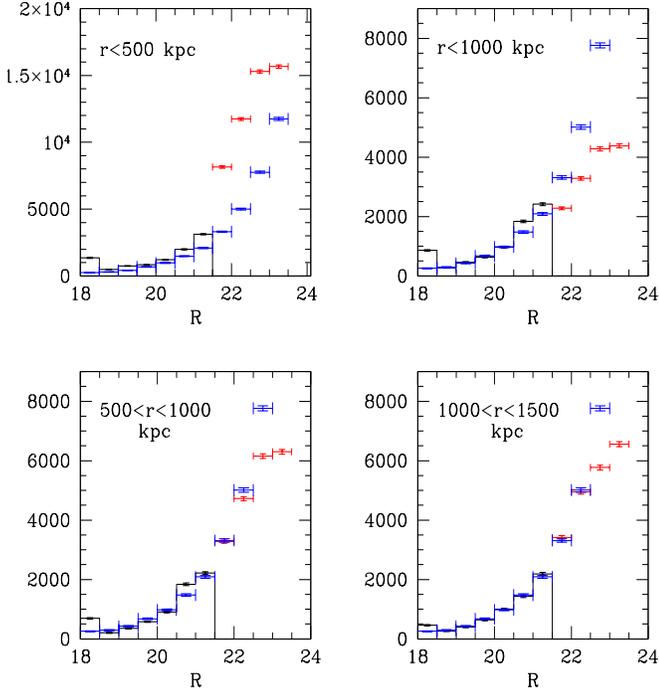,width=9cm,height=10cm,angle=0}}
\caption[]{Galaxy counts in four regions defined by their distance r to
the cluster centre in kpc. Top left: circle of radius r$<$500~kpc, top
right: circle of radius r$<$1000~kpc, bottom left: ring between 500 and
1000~kpc, bottom right: ring between 1000 and 1500~kpc. The black
histograms correspond to galaxies selected as in
Fig.~\ref{fig:sep_gal_star} and red histograms correspond to galaxy
counts estimated statistically.  The blue histograms show the VVDS counts
as described in Section~\ref{sec:VVDS}. Error bars are Poissonian.
All counts are normalized to 1~deg$^2$.}
\label{fig:gal_counts}
\end{figure}

Galaxy counts were made in four regions defined by their distance r to
the cluster centre in kpc: a circle of radius r$<$500~kpc, a circle of
radius r$<$1000~kpc, a ring between 500 and 1000~kpc, and a ring between
1000 and 1500~kpc. Note that except for the first, all of these regions
are truncated since our images do not fully cover them (see
Fig.~\ref{fig:map}). The areas covered by each of these four
regions after masking bright stars are given in Table~\ref{tab:area}.
Galaxy counts are displayed in Fig.~\ref{fig:gal_counts}. Error bars are
Poissonian.

\subsection{Comparison field}
\label{sec:VVDS}

In order to perform a statistical subtraction of the background
contribution, we consider the galaxy counts in the F02 VVDS
field\footnote{http://cencosw.oamp.fr/VVDSphot/F02}, which was
observed with the same filters as Abell~780, and is prone to little
extinction since it is located at high galactic latitude
(58.0$^\circ$). It covers a total area of 1.18 deg$^2$ and is fully
complete and free from surface brightness selection effects up to
${\rm I_{AB}=24}$, with more than 50\% completeness at B=26.5 and
R=25.9 (see McCracken et al. 2003, Section~4.1).

We will take the galaxy counts in 0.5 magnitude intervals normalized to
1~deg$^2$ directly from Table~2 of McCracken et al. (2003). Since our
magnitudes were transformed to the AB system, the comparison of the
galaxy counts is straightforward. The VVDS galaxy counts are shown in
Fig.~\ref{fig:gal_counts}, where all counts are
normalized to 1~deg$^2$. We can see from this
figure that it is only in the r$<$500~kpc region that the cluster galaxy
counts are notably above the background counts up to R$\sim 23.5$. In
the 500$<$r$<$1000~kpc and 1000$<$r$<$1500~kpc regions, there is overall
agreement between the galaxy counts and those of the VVDS, implying that
the cluster does not show strongly above the background for r$>$500~kpc.
This explains why the galaxy counts in a r$<$1000~kpc circle are so
strongly diluted that they are not significantly higher than the
background counts.

We will therefore only analyze the galaxy luminosity function in the
r$<$500~kpc region hereafter.

\section{Results: colour-magnitude diagram and galaxy luminosity functions}

\subsection{Colour--magnitude diagram}

\begin{figure} 
\centering \mbox{\psfig{figure=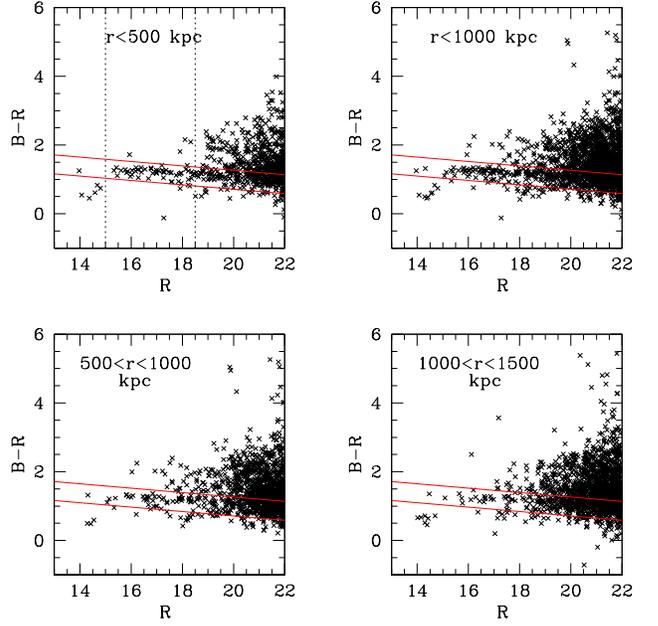,width=9cm}}
\caption[]{B$-$R vs. R colour--magnitude diagram in the same regions
  as Fig.~\ref{fig:gal_counts}. The red lines show the interval in
  which galaxies are considered as belonging to Abell 780. The
  vertical dotted lines on the top left figure indicate the magnitude
  interval in which the colour-magnitude relation was
  computed.} \label{fig:coulmag}
\end{figure}

Since the McCracken et al. (2003) galaxy counts that we will use to
subtract the background contribution start at R=18, we need a selection
criterium to build the GLFs for R$<18$. 

The B$-$R vs. R colour--magnitude diagram is shown in
Fig.~\ref{fig:coulmag} for the same four regions as in
Fig.~\ref{fig:gal_counts}. A sequence is well defined for galaxies in
the magnitude range 15$<$R$<$18.5, particularly in the plot
corresponding to a radial distance to the cluster centre smaller than
500~kpc. We computed the best fit to the B$-$R vs. R relation in this
zone and in this magnitude range by applying a simple linear
regression. We then eliminated the galaxies located at more than
3$\sigma$ from this relation and computed the B$-$R vs. R relation
again.  The equation of the colour--magnitude relation is found to be:
B$-$R=$-0.064$R+2.27 with a $\pm 2\sigma$ scatter of $\pm 0.28$. We
will consider that all the galaxies located within 2$\sigma$ of this
relation (i.e. between the two red lines of Fig.~\ref{fig:coulmag})
belong to the cluster. An eye check of the 5 very bright objects
  (R$<15$) below the sequence shows that 3 are actually stars with
  small diffraction crosses (misclassified by SExtractor as galaxies),
  one is the blend of a bright galaxy plus a star and one is a star with one
  or two objects superimposed.

\subsection{Global galaxy luminosity functions}

\begin{table}
\caption{Galaxy counts (in 0.5 mag bins) and luminosity functions in the
 B and R bands in a region within 500~kpc of the cluster centre,
 normalized to a 1~deg$^2$ area (numbers are not given when the GLF
 is negative).}
\begin{center}
\begin{tabular}{|rrr|rrr|}
\hline
 B     &  NB$_{nor}$ & GLF$_{B}$& R  &    NR$_{nor}$ & GLF$_{R}$ \\
\hline
13.75  &  &       0   &    13.75 & &     20 \\
14.25  &  &       0   &    14.25 & &      0 \\
14.75  &  &       0   &    14.75 & &      0 \\
15.25  &  &      20   &    15.25 & &     61 \\
15.75  &  &       0   &    15.75 & &     101 \\
16.25  &  &       0   &    16.25 & &     162 \\
16.75  &  &      81   &    16.75 & &     162 \\
17.25  &  &     162   &    17.25 & &     203 \\
17.75  &  &     122   &    17.75 & &     223 \\
18.25  &  &     264   &    18.25 & &     203 \\
18.75  &     223  &    157  &    18.75 &      487  &    192  \\
19.25  &     223  &    123  &    19.25 &      750  &    323  \\
19.75  &     304  &    156  &    19.75 &      831  &    155  \\
20.25  &     527  &    252  &    20.25 &     1217  &    240  \\
20.75  &     487  &     60  &    20.75 &     1987  &    508  \\
21.25  &    1217  &    476  &    21.25 &     3122  &   1033  \\
21.75  &          &         &    21.75 &     8148  &   4837  \\
22.25  &          &         &    22.25 &    11743  &   6731  \\
22.75  &          &         &    22.25 &    15296  &   7534  \\
\hline
\end{tabular} \label{tab:fdl}
\end{center}
\end{table}

\begin{figure} 
\centering \mbox{\psfig{figure=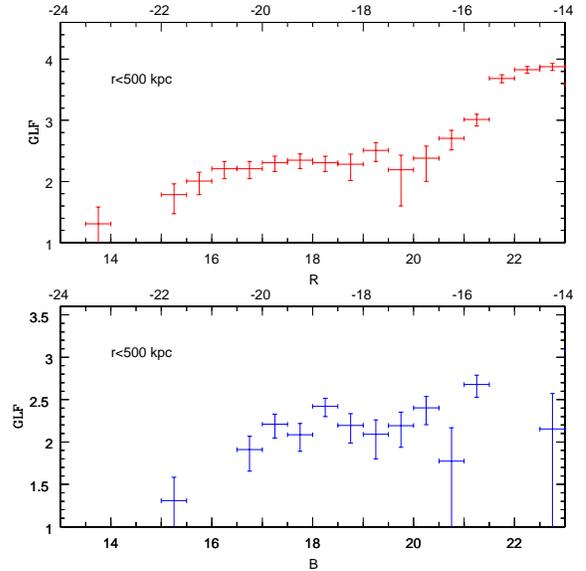,width=8cm}}
\caption[]{Galaxy luminosity functions in a 500~kpc radius region around
the cluster centre in the R (top) and B (bottom) bands, normalized to
1~deg$^2$, in logarithmic scale.  Apparent and absolute magnitudes are
indicated in the bottom and top scales respectively.}
\label{fig:fdl_all}
\end{figure}

Galaxies belonging to Abell~780 with magnitudes R$\leq 18.5$ were
selected by applying the colour--magnitude selection described above.
This limit of 18.5 was chosen rather than the lower magnitude limit
of 18.0 of the VVDS counts in order to avoid possible errors due to
small number statistics in the very first VVDS bin.  Numbers of
galaxies were computed in bins of 0.5 magnitude.

For galaxies with magnitudes $18.5<R<23$, GLFs were estimated in bins of
0.5 magnitude by simply subtracting the VVDS F02 counts from the
Abell~780 counts in the B and R bands, estimated as described in
Section~2.6. All counts were normalized to a 1~deg$^2$ region. Results
are given in Table~\ref{tab:fdl} and Fig.~\ref{fig:fdl_all}
for the region of 500~kpc radius
around the cluster center.  For magnitudes above B=21.5 (except for a
point at B=22.75, which has a very large error bar, as it is also the
case for the point at B=20.75) and R$>$23, our galaxy counts become
smaller than those in the background.

The error bars drawn in Fig.~\ref{fig:fdl_all} were taken to be 4
times the Poissonian errors on galaxy counts, as suggested by detailed
simulations previously performed by our team (see Bou\'e et al. 2008,
Fig.~5).

We can see,
particularly in R where the signal to noise ratio is better and the GLF
smoother, that the GLF is composite, with a broadly shaped distribution
at bright magnitudes, and a rising part for R$>$19.5.  A
power law fit between R=20.25 and R=21.75 (corresponding to absolute
magnitudes between $-16.65$ and $-15.15$) performed with a linear
regression gives a slope $\alpha = -0.85\pm 0.12$ (when expressed as a
function of absolute R magnitude). 

The bright part of the GLF (R$< 18$, or ${\rm M_R<-18.9}$) is very
similar in shape to the GLFs recently obtained e.g. by Andreon et
al. (2008), who analyzed the GLFs of a sample of clusters at various
redshifts, limited to bright absolute magnitudes ${\rm M_V}<-19$.

The faint end slope is somewhat flatter than usually found in
clusters, where it is usually between $-1.0$ and $-2.0$ in the same
absolute magnitude range (see e.g. Bou\'e et al. 2008, Table
A.1). Besides, it is somewhat surprising that for R$>$22 the GLF
  flattens since the completeness was estimated to be better than 90\%
  uo to R=23.  Note however that the region that we consider is
rather small (500~kpc in radius, corresponding to 0.26r$_{200}$) and
corresponds to the cluster centre, where the slope is expected to be
flatter. This would agree with the general picture of a relaxed
cluster where the central cD galaxy has had time to accrete many
surrounding dwarf galaxies without being refueled by numerous infalls
from the field, thus leading to a number of faint galaxies smaller
than expected.  However, Fig.~\ref{fig:gal_counts} seems to show
  that, if there are any dwarfs left, then they are in the centre, in
  contradiction with the previous explanation.  
  One possible reason (qualitatively supported for example by Adami et
  al. 2007b) could be that faint galaxies in clusters (i.e. here
  R$>20$ or ${\rm M_R}>-17$) could have different origins. Moderately
  faint galaxies (20$<$R$<$22) could be classically accreted faint
  field galaxies, while fainter objects (R$>22$) could be cluster-made
  galaxies, or in other words debris of larger galaxies. In this
  picture, we can explain the rising GLF of Abell~780 up to R$\sim$22
  as being driven by normal accreted field galaxies. The flatter GLF
  at R$>$22 would then be explained by the quite isolated and quiet
  nature of A780, likely to reduce interaction processes and therefore
  to inhibit the creation of faint debris inside the cluster. However,
  this scenario, proposed for the Coma cluster, may not be valid here,
  since it does not explain why there are fewer moderately faint
  galaxies in the region between 500 and 1000~kpc than in the
  innermost 500~kpc radius region.

A comparison of the R band GLF of Abell~780 (in the region of 500~kpc
radius) with the GLFs in the North and South halves of the Coma cluster
(Adami et al. 2007a) shows that the GLF shapes are quite similar up to
the absolute magnitude ${\rm M_R} \sim -14.5$. However, our data
are not as deep as those of Coma, so we cannot compare the very faint
ends, corresponding to ${\rm M_R} > -14.5$ (where the Coma North and
South GLFs differ, due to the influence of the environment at large
scale).

The GLF in the B band does not show any obvious difference with that in
the R band; since it is much noisier, we will not discuss it
further.

\subsection{Galaxy luminosity functions in various regions}

\begin{figure} 
\centering
\mbox{\psfig{figure=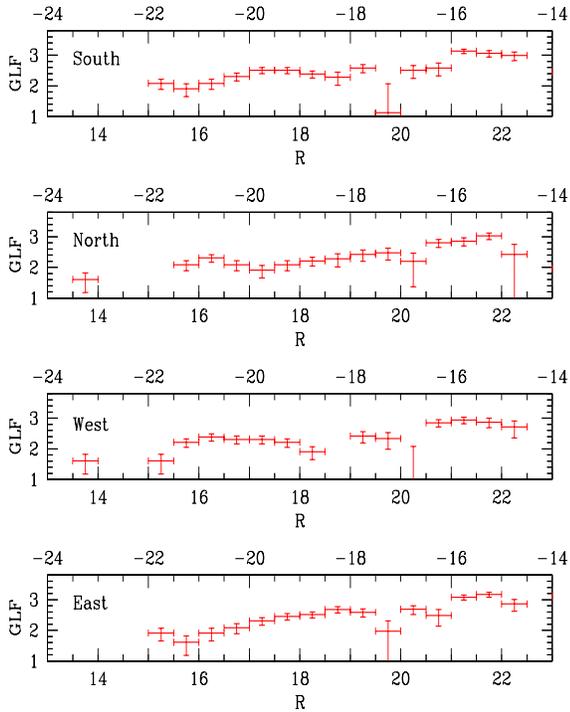,width=8cm,height=10cm}}
\caption[]{Galaxy luminosity functions in the R band normalized to an
  area of 1~deg$^2$. The 500~kpc radius region around the cluster
  centre was divided in halves: from top to bottom South half, North
  half, West half and East half. }
\label{fig:fdl_44}
\end{figure}

In order to see if any difference suggesting environmental effects could
be found, we analyzed the GLF in various regions of the cluster. For
this, we cut the 500~kpc radius region into two halves: one North and
one South of the cluster centre, then two other halves: one East and one
West of the cluster centre. The GLFs were derived as described above in
those four regions, in the R band only since it is the band with the
highest number of galaxy counts. The GLFs obtained are given in
Fig.~\ref{fig:fdl_44} (normalized to an area of 1~deg$^2$).  We can
see that in the West and South zones their shapes are more similar to
the overall cluster GLF than in the East and North where the GLFs rise
more or less continuously, but since they are rather noisy it is
difficult to say much more.


Altogether, environmental effects appear rather weak in Abell~780. We
describe below a short analysis of the Abell~780 cluster environment,
performed to look for apparent substructures such as filaments along
which galaxy infall could be occuring.

\section{Discussion and conclusions}

\subsection{The central regions of the cluster: comparison with X-ray data}

\begin{figure} 
\centering
\mbox{\psfig{figure=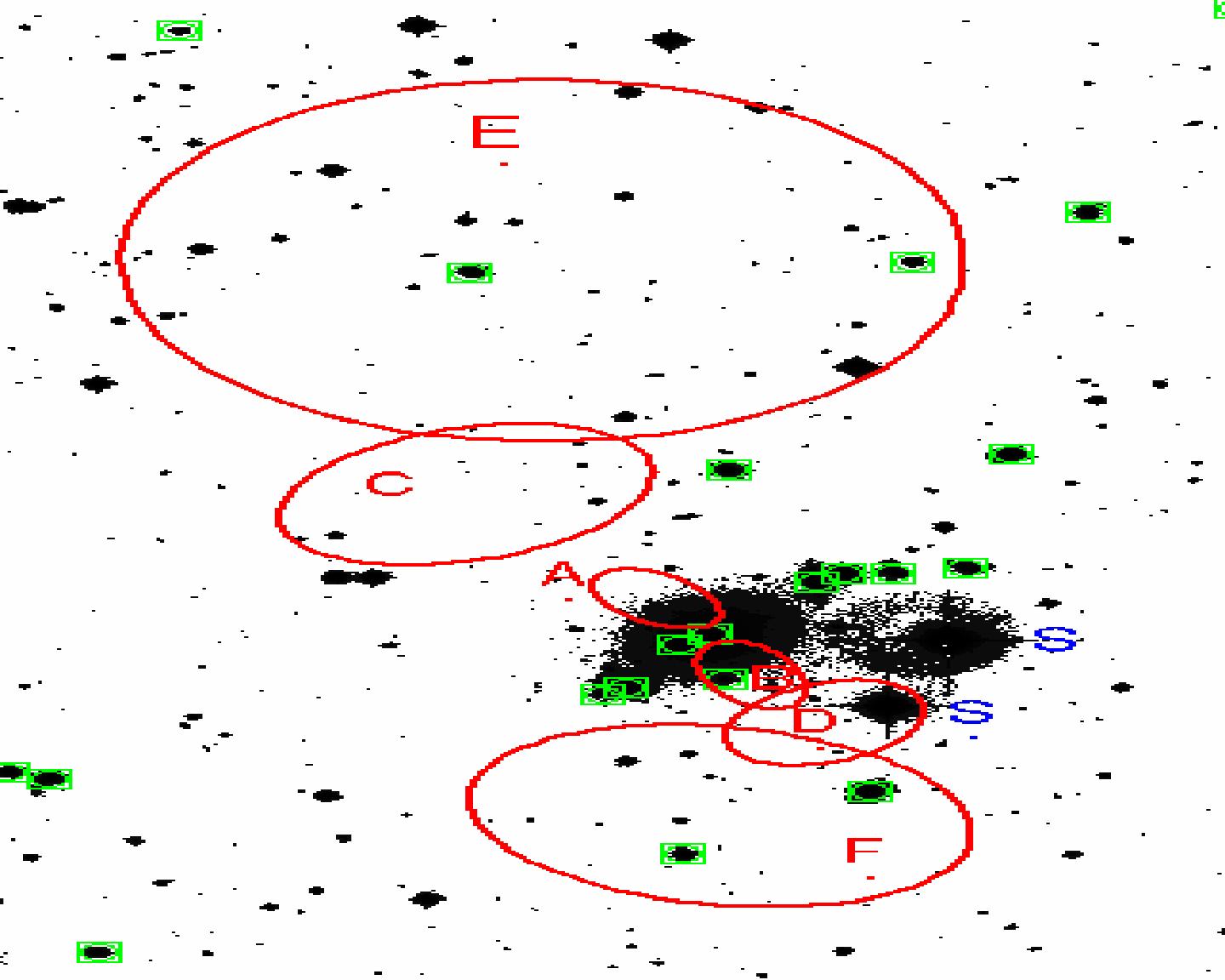,width=5.5cm,height=9cm}}
\caption[]{Inverted colour R band image of the cluster centre with the
positions of the X-ray cavities found by Wise et al. (2007) superimposed
and labeled in red.  The S symbol shows the positions of the two bright
stars close to the cD.  Galaxies brighter than R=18 are indicated with a
green square. North is top and East is left.}  \label{fig:mcnamara}
\end{figure}

Abell~780 is well known in X-rays under the name of Hydra~A,
originating from the fact that the cluster cD is a strong radio
emitting galaxy bearing this name. Based on a deep Chandra image, Wise
et al. (2007) found several X-ray cavities (labeled A to F) in the
central region, which they explained by outflow (continuous or in
bursts) from the nucleus of the central galaxy. The cavities could
each harbor one or several AGN (expected to be intrinsically bright
and to have bluer colours than normal galaxies).  The optical R band
image around the cluster centre is shown in Fig.~\ref{fig:mcnamara}
with the positions of the X-ray cavities from Wise et
al. superimposed. Cavities A and B are partly superimposed on the cD,
and SExtractor detects several galaxies brighter than R=18
superimposed on the cD (but slightly outside A and B). No bright
galaxy is visible in cavities C and D. Cavity E contains two objects
brighter than R=18, but one of them is probably a star (a weak
diffraction cross is visible when carefully looking at the
image). Cavity F has two bright galaxies with R magnitudes 16.2 and
16.8; the brighter one has a redshift z=0.051526 according to NED, but
all three galaxies in cavities E and F have normal B-R colours, so
either spectroscopy or high resolution multiband imaging (to search
for blue cores) are necessary to check if they are AGN; to our
knowledge they have no X-ray or radio counterparts.

\subsection{Large scale structure}

\begin{figure} \centering
\mbox{\psfig{figure=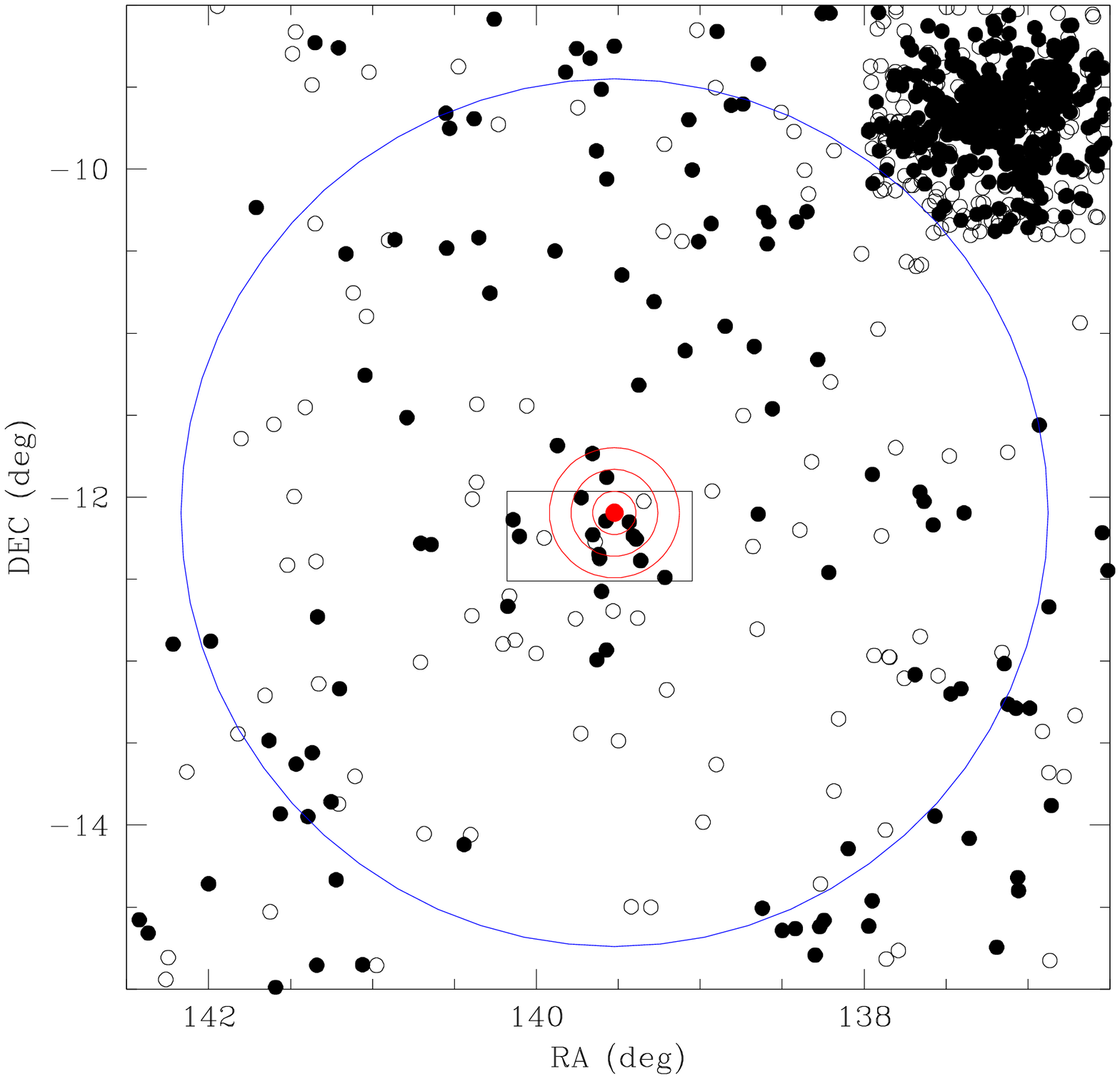,width=7cm,height=7cm,angle=0}}
\mbox{\psfig{figure=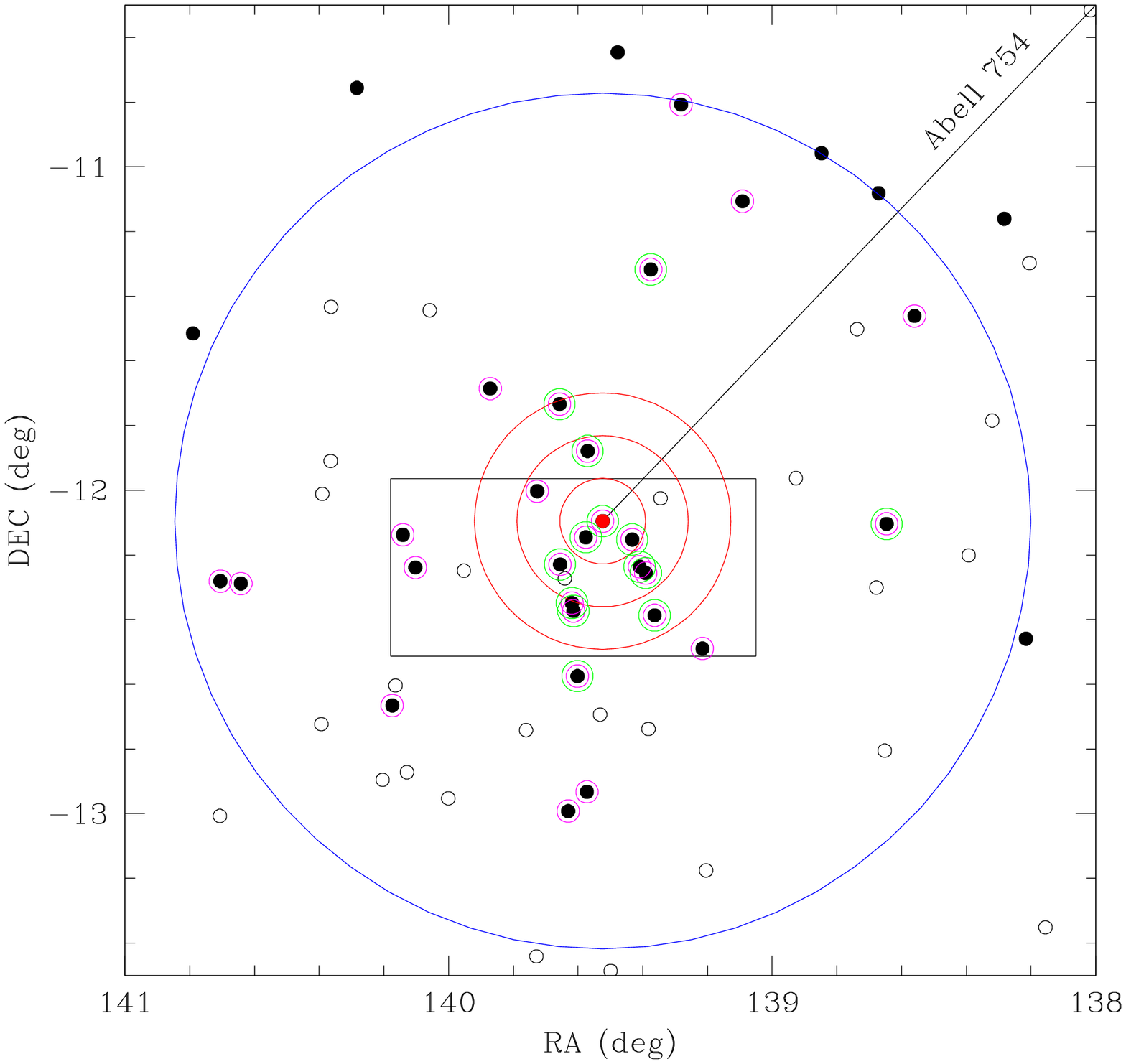,width=7cm,height=7cm,angle=0}}
\caption[]{Positions of the galaxies with measured redshifts and
magnitudes in a large scale region around Abell 780, taken from
NED. Small filled black circles show galaxies with redshifts in the
[0.0338, 0.0738] redshift interval, while empty circles show galaxies
with redshifts outside this range.  The black rectangle indicates the
total field covered by our images. The red filled square corresponds to
the cluster centre as defined in the text. The three small red
concentric circles correspond to radii of 0.5, 1.0 and 1.5~Mpc. Top
figure: the large blue circle corresponds to a radius of 10.0~Mpc.  The
Abell~754 cluster is visible at the top right. Bottom figure: zoom of
the top figure, with the large blue circle having a radius 5.0~Mpc.  The
27 galaxies circled in magenta and the 14 galaxies circled in green
correspond to the two gravitationally bound
structures according to the Serna--Gerbal analysis. The line indicates
the direction towards Abell~754.}  \label{fig:large_scale}
\end{figure}

Since clusters are believed to form from material falling along
filaments, environmental effects are expected to have an important
influence on their properties. We have searched the NED database in the
environment of Abell~780 and found 1219 galaxies with measured redshifts
in a region of 5~deg (18.9~Mpc) radius around the cluster. Out of these,
719 have redshifts in the [0.0338,0.0738] range, i.e. within $\pm
6000$km~s$^{-1}$ of the mean velocity of Abell~780, a broad
velocity interval corresponding to $\pm 8\sigma _v$, chosen to avoid
``losing'' galaxies that could be linked to Abell~780.  In order to
search for gravitationally bound structures, we selected a subsample of
galaxies which also had magnitudes available; the spatial distribution
of these galaxies is shown in Fig.~\ref{fig:large_scale}. The Abell~754
cluster is clearly visible North-West of Abell~780, at a distance of
about 12.8~Mpc; its redshift is 0.0542, almost the same as that of
Abell~780.  However, no galaxy filament is detected between Abell~780 and
Abell~754.

In two smaller regions of radii 10~Mpc and 5~Mpc around Abell~780,
there are 202 and 85 galaxies with measured redshifts respectively
(not all with measured magnitudes), with 107 and 58 galaxies in the
[0.0338,0.0738] redshift range respectively. If we consider galaxies
in this redshift range and cut the 5~Mpc circle in halves, we find
that there are 41, 17, 29 and 29 galaxies in the south, north, west
and east halves respectively. If we divide the cluster in halves, with
the dividing line oriented perpendicular to the line connecting
Abell~780 to Abell~754, we find fewer galaxies in the half circle
closer to Abell~754 than in the other one (23 against 35).  These
numbers are too small for statistics, and we have no information on
the NED data coverage and completeness, so all we can say is that,
according to these data, there is no evidence for a galaxy
density enhancement towards Abell~754, on the contrary there may be
more galaxies south of the cluster at large scale.

\subsection{Search for gravitationally bound structures}

In order to search for gravitationally bound structures, we applied
the Serna-Gerbal (1996, hereafter SG) method to the redshift and
magnitude catalogue that we compiled within a radius of 5~Mpc (see
above). The SG method allows gravitationally bound galaxy subgroups to
be extracted from a catalogue containing positions, magnitudes, and
redshifts, based on the calculation of their relative (negative)
binding energies.  This calculation takes into account the mass to
luminosity M/L ratio chosen by the user as an input value, but the
group mass derived later is estimated from the group binding energy
and velocity dispersion, and does not depend upon M/L, which only acts
as a contrast criterium.  We will assume here M/L=400.  Qualitatively,
high input values of M/L only allow the detection of the major
structures, while low values of M/L allow to detect structures with
low link energy. Since we only have a limited redshift and magnitude
catalogue, we will not search for small structures in detail. In any
case, results do not depend strongly on the value of M/L, since a
variation of a factor of two in this parameter does not change
significantly the results.  Note however that our redshift catalogue
is extracted from NED, and since we do not have completeness
information, we cannot estimate quantitatively how strong the results
of the SG analysis are.

The SG method gives as output a list of galaxies belonging to each
group, as well as information on the binding energy and mass of the
group itself, with a minimum number of 3 galaxies per group.


\begin{figure} 
\centering
\mbox{\psfig{figure=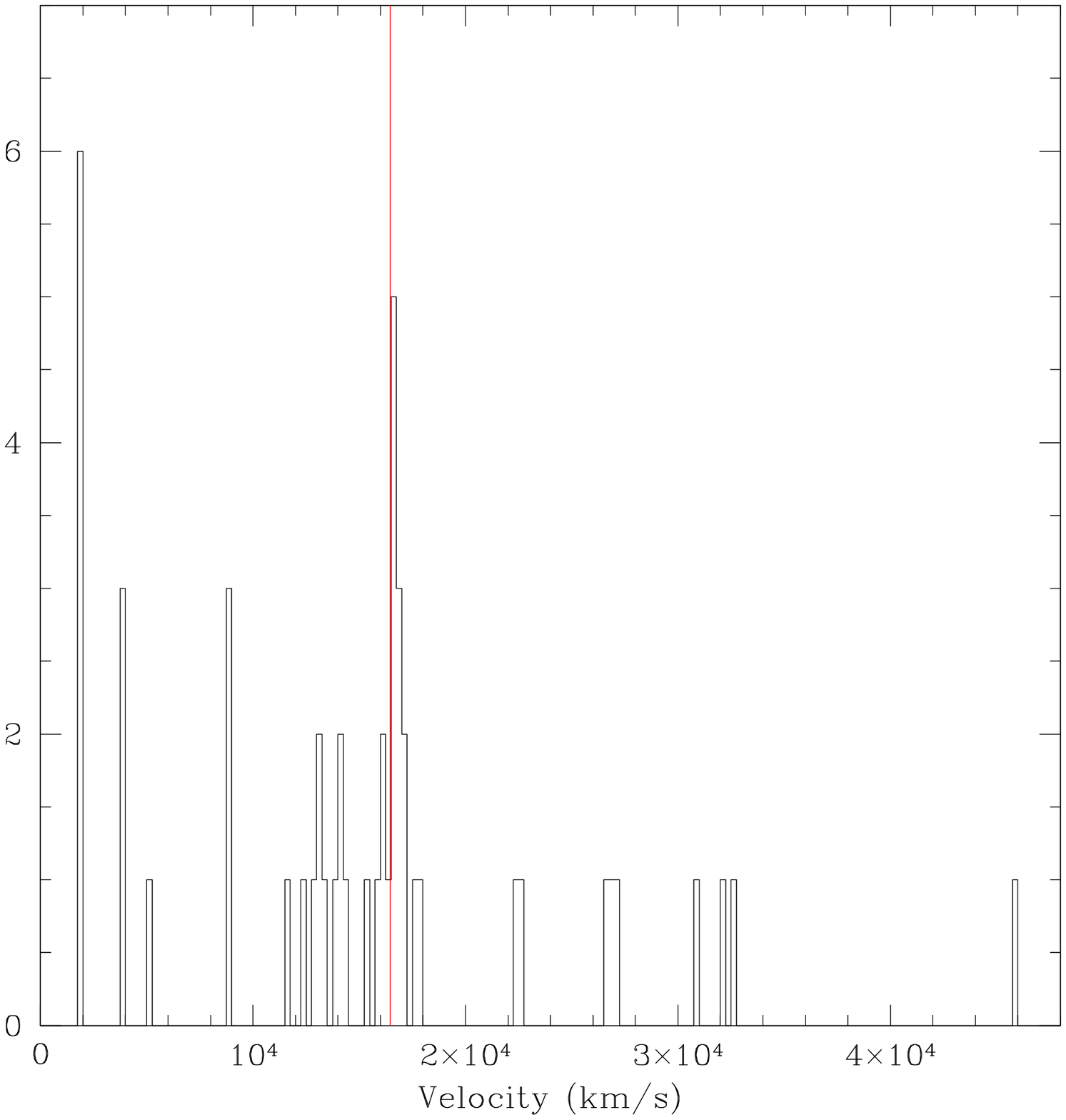,width=6cm,height=6cm,angle=0}}
\mbox{\psfig{figure=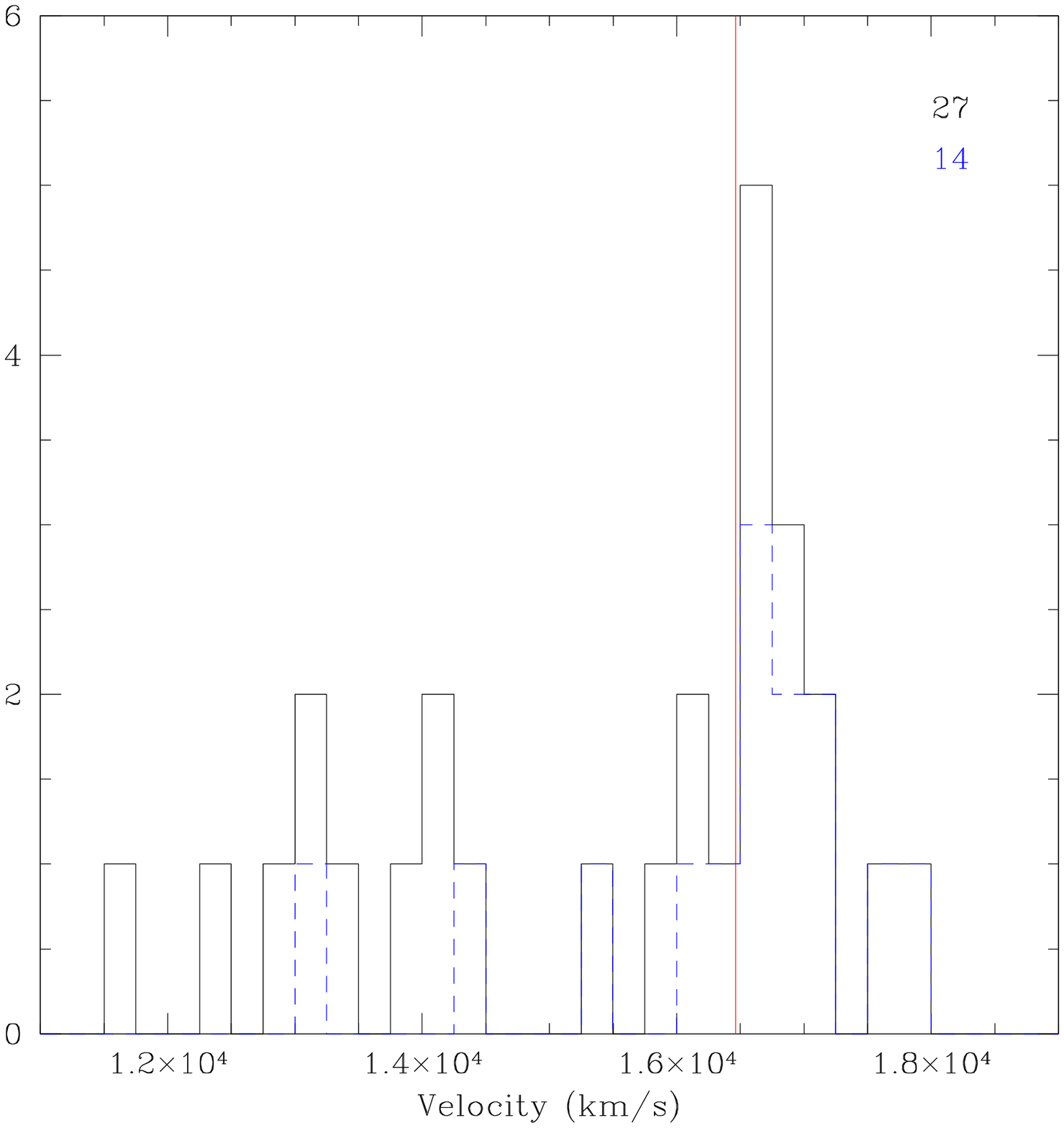,width=6cm,height=6cm,angle=0}}
\caption[]{Top: velocity histogram of the 49 galaxies with measured
  redshifts and magnitudes in a 5~Mpc circle around Abell~780. Bottom:
  velocity histograms of the 27 (in black) and 14 (in blue) galaxies
  found to be gravitationally bound by the SG method.  The red
  vertical line indicates the velocity of the cD galaxy of Abell~780.}
\label{fig:histovit_21}
\end{figure}

Within a 5~Mpc radius, once galaxies with no magnitudes available are
excluded, we are left with a catalogue of 49 galaxies with magnitudes
brighter than 16. Out of these 49 galaxies, 16 have no filter
indication in NED, 21 have magnitudes in the R filter and 12 in the I
filter. In order to make magnitudes as homogeneous as possible, we
converted the I magnitudes into R magnitudes assuming R-I=0.7, taken
for a typical elliptical galaxy at z=0 from Fukugita et al. (1995).

The SG analysis of this catalogue shows the existence of a
gravitationally bound structure of 27 galaxies (circled in magenta in
Fig.~\ref{fig:large_scale}).  The velocity histogram of these 27
galaxies is displayed in Fig.~\ref{fig:histovit_21} (bottom, black
histogram), together with the velocity histogram of the full sample of
49 galaxies (top).  The biweight velocity dispersion of these 27
galaxies is 1968~km~s$^{-1}$, a value notably higher than the overall
cluster velocity dispersion of $\sigma _v$ = 758 km~s$^{-1}$ estimated
in section~1 and suggesting that substructuring remains within these
27 galaxies, as also suggested by the velocity histogram of
Fig.~\ref{fig:histovit_21}. It is likely that out of these 27
galaxies, 10 are probably foreground objects.  The mass of $4.4\
10^{13}$~M$_\odot$ estimated for the 27-galaxy structure is therefore
most probably an overestimate. If we exclude the 10 foreground objects
and consider only the 17 galaxies which are likely to be
gravitationally bound, their velocity dispersion is then 583~km/s.


A more strongly gravitationally bound structure of 14 galaxies is
found within the above structure of 27 galaxies. It has a mass of
$1.5\ 10^{13}$~M$_\odot$ and a velocity dispersion of 875~km~s$^{-1}$.
Its velocity histogram is shown in blue in Fig.~\ref{fig:histovit_21}
(bottom) and suggests that 2 galaxies are foreground objects; if these
2 galaxies are excluded, the velocity dispersion becomes
640~km~s$^{-1}$.

If we now consider all the galaxies within a radius of 5~Mpc with
measured redshifts, whether they have measured magnitudes or not, we
have a list of 86 galaxies. Out of these, 43 galaxies have velocities
between 14700 and 18000~km~s$^{-1}$, corresponding to the cluster
velocity interval, and the velocity dispersion of these 43 galaxies is
$\sigma _v$=741~km~s$^{-1}$, almost exactly that derived from X-rays.

In view of the optical data available for Abell~780, this cluster
appears to have an overall relaxed nature, with no or hardly any
infall of field dwarf/faint galaxies. This agrees with the large scale
X-ray properties of Abell~780 derived from XMM-Newton data by
Simionescu et al.  (2009a, 2009b). However, deeper and broader field
imaging and more galaxy redshifts are obviously necessary to achieve a
better understanding of the Abell~780 cluster, particularly at small
scales where Chandra X-ray data have revealed a wealth of structures
(Wise et al. 2007).

\begin{acknowledgements}

We warmly thank the referee for many constructive comments which helped
 improve the paper. We acknowledge help from G.~Mars and G.~Daste for
 part of the data reduction.  We are very grateful to G.~Bou\'e for
 making several of his softwares available to us and for helping us to
 use them. We also thank Brian McNamara for sending us the ds9 region
 file showing the positions of the X-ray cavities.

\end{acknowledgements}


\begin{thebibliography}{}
                             
\bibitem{} Adami C., Picat J.-P., Savine C. et al. 2006a, A\&A 451, 1159

\bibitem{} Adami C., Scheidegger R., Ulmer M. et al. 2006b, A\&A 459,
	679

\bibitem{} Adami C., Durret F., Mazure A. et al. 2007a, A\&A 462, 411

\bibitem{} Adami C., Picat J.P., Durret F. et al. 2007b, A\&A 472, 749

\bibitem{} Andreon S., Puddu E., de Propris R., Cuillandre J.-C. 2008,
MNRAS 385, 979
	
\bibitem{} Batuski D.J., Burns J.O., Newberry M.V. et al. 1991, AJ 101, 1983

\bibitem{} Beijersbergen M., Hoekstra H, van Dokkum P.G., van der Hulst T.
 2002, MNRAS 329, 385

\bibitem{} Bertin E. \& Arnouts S. 1996, A\&AS 117, 393

\bibitem{} Bou\'e G., Adami C., Durret F., Mamon G., Cayatte V. 2008, 
	 A\&A 479, 335


\bibitem{} Burns J.O., Rhee G., Owen F.N., Pinkney J. 1994, ApJ 423, 94

\bibitem{} Carlberg R.G., Yee H.K.C., Ellingson E. 1997, ApJ 478, 462

\bibitem{} David L.P., Arnaud K.A., Forman W., Jones C. 1990, ApJ 356, 32

\bibitem{} Fasano G., Marmo C., Varela J. et al. 2006, A\&A 445, 805

\bibitem{} Fukugita M., Shimasaku K., Ichikawa T. 1995, PASP 107, 945

\bibitem{} Girardi M., Fadda D., Giuricin G. et al. 1996, ApJ 457, 61

\bibitem{} Lobo C., Biviano A., Durret F. et al. 1997, A\&A 317, 385	

\bibitem{} McCracken H.J., Radovich M., Bertin E. et al. 2003, A\&A
	410,17

\bibitem{} McNamara B.R., Wise M., Nulsen P.E.J. et al. 2000, ApJ 534, L135

\bibitem{} Mohr J.J., Mathiesen B., Evrard A.E., 1999, ApJ 517, 627

\bibitem{} Owen F.N., White R.A., Burns J.O. 1992, ApJS 80, 501

\bibitem{} Ramella M., Biviano A., Pisani A. et al. 2007, A\&A 470, 39

\bibitem{} Robin A.C., Reyl\'e C., Derri\`ere S., Picaud S. 2003,
A\&A 409, 523

\bibitem{} Schlegel D.J., Finkbeiner D.P., Davis M. 1998, ApJ 500, 525

\bibitem[]{} Serna A. \& Gerbal D. 1996, A\&A 309, 65

\bibitem{} Simionescu A., Werner N., B\"ohringer H. et al. 2009a, A\&A
	493, 409

\bibitem{} Simionescu A., Roediger E., Nulsen P.E.J. et al. 2009b, A\&A
	495, 721

\bibitem{} Wise M.W., McNamara B.R., Nulsen P.E.J., Houck J.C., David L.P. 2007,
ApJ 659, 1153

\end{thebibliography}
\end{document}